\documentclass[a4paper,fleqn,usenatbib]{mnras}

\usepackage{savesym} 

\usepackage[T1]{fontenc}
\usepackage{ae,aecompl}

\usepackage{natbib}
\usepackage{graphicx}	
\usepackage{amsmath}	
\usepackage{amssymb}	

\savesymbol{iint}
\usepackage{txfonts}
\restoresymbol{TXF}{iint} 

\usepackage{soul}
\usepackage{color}
\sethlcolor{cyan}

\title{Non-Detection of Nova Shells Around Asynchronous Polars}

\author[Pagnotta \& Zurek]{
Ashley Pagnotta$^{1}$\thanks{E-mail: pagnotta@amnh.org}
and David Zurek$^{1}$
\\
$^{1}$Department of Astrophysics, American Museum of Natural History, New York, NY 10024
}

\date{Accepted 2016 February 19. Received 2016 February 5; in original form 2016 February 5}

\pubyear{2016}

\begin{document}
\label{firstpage}
\pagerange{\pageref{firstpage}--\pageref{lastpage}}
\maketitle

\begin{abstract}
Asynchronous polars (APs) are accreting white dwarfs (WDs) that have different WD and orbital angular velocities, unlike the rest of the known polars, which rotate synchronously (i.e., their WD and orbital angular velocities are the same). Past nova eruptions are the predicted cause of the asynchronicity, in part due to the fact that one of the APs, V1500 Cyg, was observed to undergo a nova eruption in 1975. We used the Southern African Large Telescope 10m class telescope and the MDM 2.4m Hiltner telescope to search for nova shells around three of the remaining four APs (V1432 Aql, BY Cam, and CD Ind) as well as one Intermediate Polar with a high asynchronicity (EX Hya). We found no evidence of nova shells in any of our images. We therefore cannot say that any of the systems besides V1500 Cyg had nova eruptions, but because not all post-nova systems have detectable shells, we also cannot exclude the possibility of a nova eruption occurring in any of these systems and knocking the rotation out of sync.
\end{abstract}


\section{Introduction}
\label{sec:intro}

Polars are accreting white dwarf (WD) binaries characterized by the presence of a strong magnetic field ($\sim 7-230$ MG; \citealp{ferrario2015a}), which prevents the formation of an accretion disc and instead channels the accreted material directly on to the poles of the WD. In most polars, the measured WD and orbital angular velocities are found to be identical. For a handful of systems, however, the WD rotates asynchronously, so that those two velocity values differ significantly. The five known asynchronous polars (APs) are V1432 Aql (RXJ 1940-10), BY Cam, V1500 Cyg, CD Ind (RXJ 2115-58), and Paloma (RX J0524+42) \citep{campbell1999a,schwarz2004a}. EX Hya is an Intermediate Polar (IP), with a magnetic field of ${\sim}1$ MG and an accretion disc that has been observed to have dwarf nova outbursts, which also has a very high asynchronicity \citep{hellier1996a}, and was included in this study as well.

The cause of this asynchronicity is theorized to be a nova eruption: accreted hydrogen builds up on the surface of the WD until reaching a critical temperature/pressure and igniting a thermonuclear runaway in the accreted layer, the ejection of which causes the WD to spin with a higher angular velocity than before the nova \citep{campbell1999a}. V1500 Cyg (Nova Cyg 1975) was observed to be an AP after its nova eruption, lending support to this theory, although its pre-eruption status is unknown. BY Cam, V1500 Cyg, and CD Ind all have a positive $\omega / \Omega$, where $\omega$ is the synodic angular velocity of the WD primary and $\Omega$ is the orbital angular velocity. $\omega / \Omega$ is then a measurement of the asynchronicity of the system. Another way to measure asynchronicity is by looking at the percent difference between the periods, defined as $\frac{P_\mathrm{orb} - P_\mathrm{spin}}{P_\mathrm{orb}}$, where $P_\mathrm{orb}$ is the orbital period of the binary system and $P_\mathrm{spin}$ is the spin period of the WD. The percent difference and  $\omega / \Omega$, along with other general properties, are listed in Table \ref{tab:summary} for each system. BY Cam, V1500 Cyg, CD Ind, and Paloma all have positive percent differences, although they range over approximately one and a half orders of magnitude. V1432 Aql, however, is under-synchronous, with a negative $\omega / \Omega$ and percent difference, which may indicate a different formation mechanism, although at this point the details of the theory are poorly understood in general, and particularly when it comes to explaining how to obtain an under-synchronous AP. 

These systems do not remain asynchronous indefinitely; instead, they likely start returning to a synchronous state quickly after being knocked out of sync. Models of this process vary, leading to a range of estimates for the time needed to return to synchronization ($t_\mathrm{sync}$) for each system, also listed in Table \ref{tab:summary}. BY Cam has the largest range, with $t_\mathrm{sync}$ estimates ranging from $250 \pm 20$ yr \citep{pavlenko2013a} to ${>}3500$ yr \citep{honeycutt2005a}. 

With $t_\mathrm{sync}$ estimated to be just a few hundreds of years for most APs, and the postulate that the asynchronicity was originally caused by a nova eruption, it is reasonable to search for nova shells around the systems and expect to find something. Detection of such a shell would eventually allow for an estimate of the date of the nova that caused the asynchronicity and thus provide another constraint on the resynchronization time-scales as well as a further clue to the cause of the asynchronicity in the first place. \cite{sahman2015a_arxiv} searched for shells around just two of our targets, V1432 Aql and BY Cam, using the Auxiliary Port on the 4.2 m William Herschel Telescope on La Palma and did not find evidence for any shells. 

\section{Observations}
We observed two APs, V1432 Aql and CD Ind, and the IP EX Hya, in H$\alpha$ ($\lambda _\mathrm{peak} = 656.6$ nm, FWHM = 10 nm) using the SALTICAM imager on the 10m class South African Large Telescope (SALT) located at the South African Astronomical Observatory, near Sutherland, South Africa \citep{buckley2006a,odonoghue2006a}. V1432 Aql was observed for a total of 3120s, divided evenly between 26 different exposures and two nights (2013 June 29 and 2013 July 11). With the same filter, CD Ind was observed on 2013 June 28 for a total of 1560s across 13 exposures, and EX Hya on 2013 July 11 for 1800s spread over 15 different exposures. (There is effectively no guide camera that can be used with SALTICAM, hence the large number of short exposures.) Additionally, we observed BY Cam in H$\alpha$ ($\lambda _\mathrm{peak} = 656.2$ nm, FWHM = 47 nm, filter borrowed from NOAO) using the OSMOS instrument (in imaging mode) on the MDM Observatory 2.4m Hiltner telescope. Our coverage of BY Cam is much shallower than for the Southern APs\textemdash we have only two 1200s exposures, for a total of 2400s of 2.4m time on the target\textemdash but we do have a much wider field of view, with a $20\arcmin$ unvignetted diameter for OSMOS as opposed to the $8\arcmin$ diameter field of view of SALTICAM.

\begin{figure*}
\centering
\includegraphics[width=\textwidth]{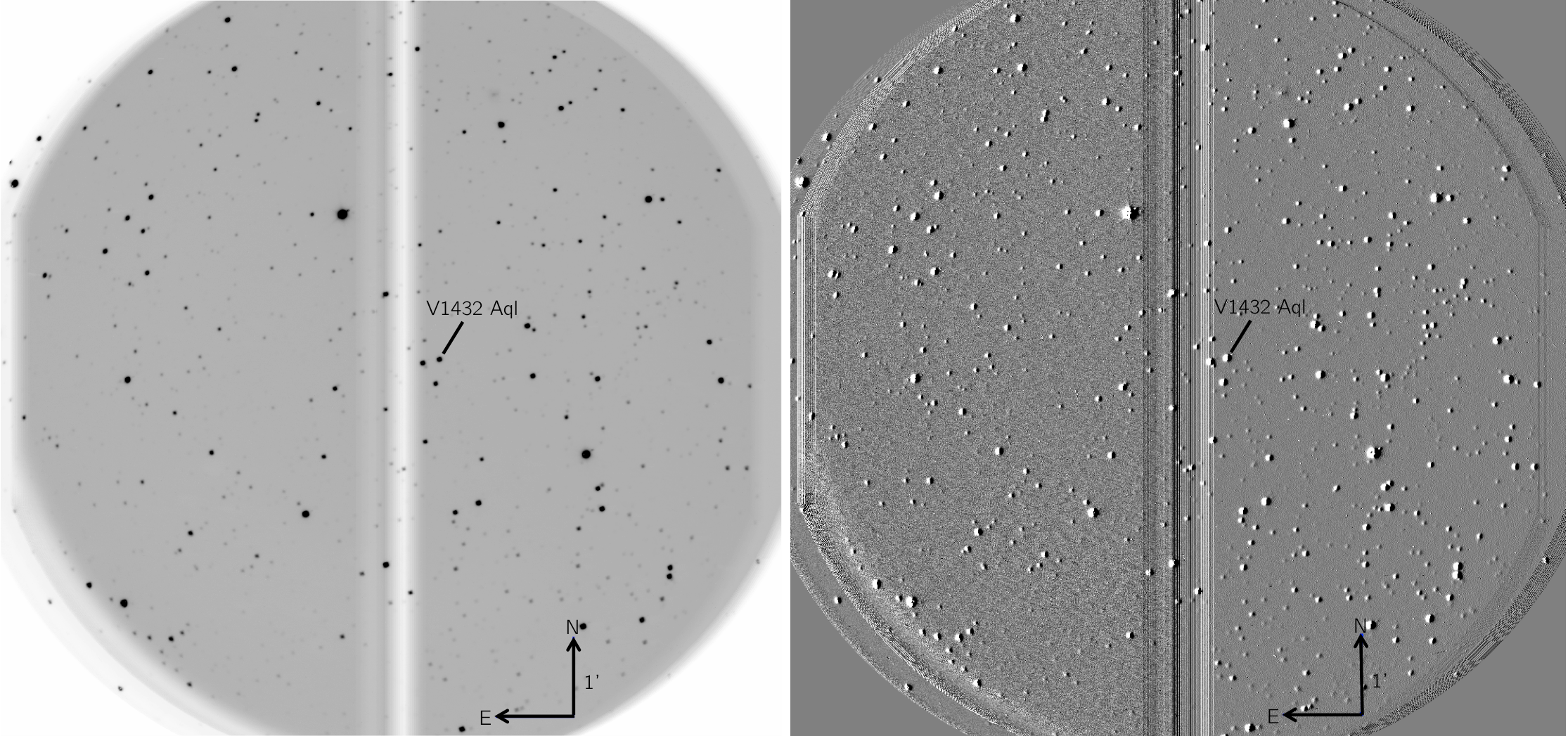}
\caption{{\it Left:} SALTICAM H$\alpha$ image of the AP V1432 Aql (RXJ 1940-10), made from a combination of 26 separate exposures taken over the course of two nights, 2013 June 29 and 2013 July 11, for a total exposure time of 3120s. North is up, East is to the left, and the length of the directional arrows corresponds to 1{\arcmin} on the figure. The full unvignetted field of view of the image is 8{\arcmin} in diameter, and the lighter stripes in the middle of the image are due to the gap between the two SALTICAM chips. The position of V1432 Aql is marked, and no shells or shell fragments are visible. {\it Right:} The right side of this figure shows the same V1432 Aql field seen on the left after it has been processed using an unsharp masking technique that involves subtracting a duplicate of the image from the original, after the duplicate has been shifted by 0.5 pixels in the x-direction. This method increases the local contrast of the different areas of the image and highlights edge features, such as those seen in nova shells. With this, we can be confident that if there were a detectable shell in this image, we would see it here.}
\label{fig:v1432aql}
\end{figure*}

\begin{figure*}
\centering
\includegraphics[width=\textwidth]{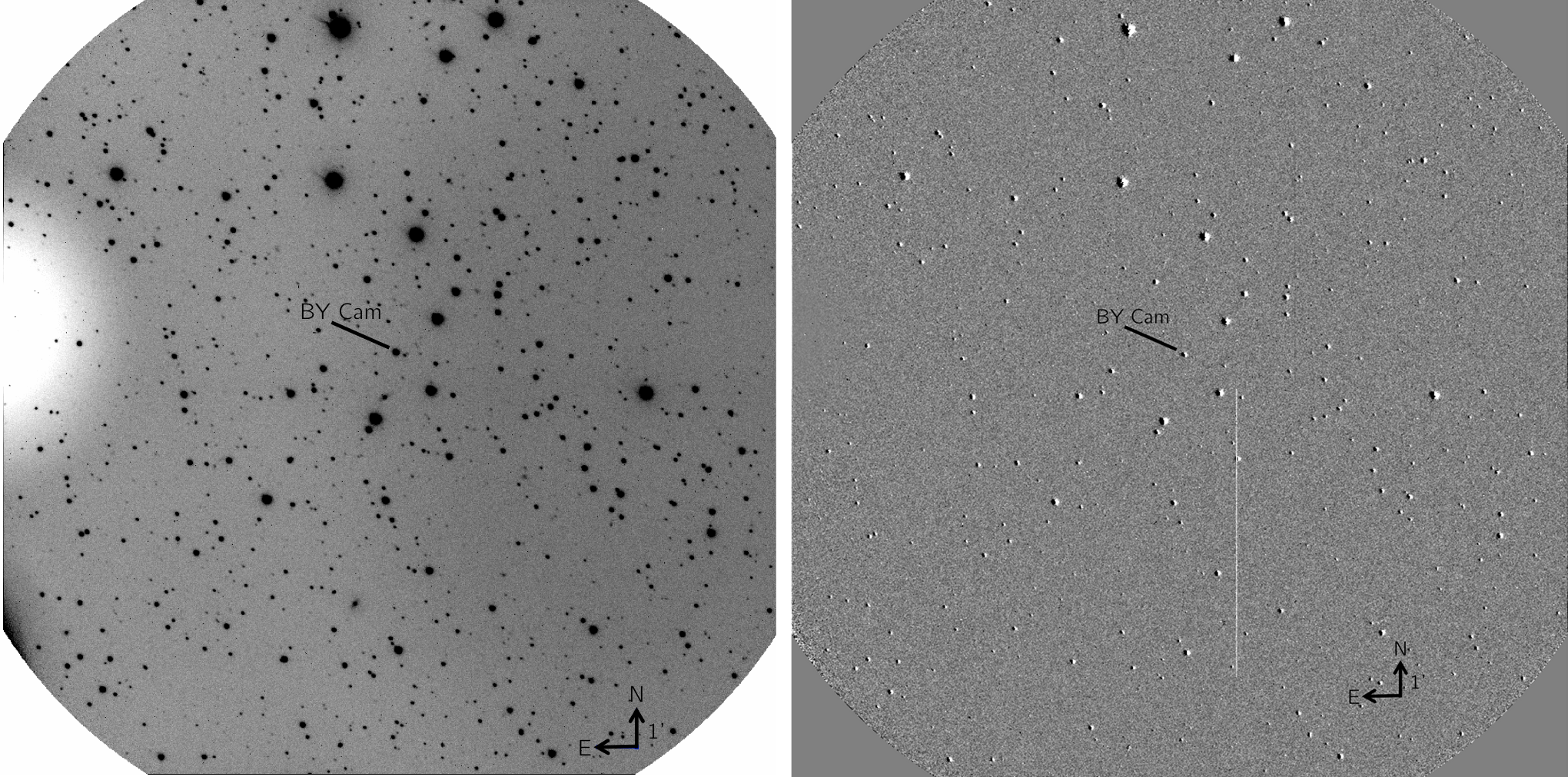}
\caption{{\it Left:} MDM 2.4m OSMOS H$\alpha$ image of the AP BY Cam, constructed from two images taken on 2015 March 15, for a total of 2400s of exposure time. North is up, East is to the left, and the length of the directional arrows corresponds to 1{\arcmin} on the figure. (The larger field of view of OSMOS compared to SALTICAM is immediately visible by comparing the relative sizes of the 1{\arcmin} lines between this figure and those in Figures \ref{fig:v1432aql}, \ref{fig:exhya}, and \ref{fig:cdind}.) The full field of view of the image is 20{\arcmin} in diameter, and the large light spot in the eastern half of the image is due to the guide camera blocking part of the frame. The position of BY Cam is marked, and no shells or shell fragments are visible. {\it Right:} This figure shows the unsharp mask technique applied to the image of BY Cam on the left side, using the same procedure described in the caption to Figure \ref{fig:v1432aql}. Again, no shells or shell fragments are visible in the image.}
\label{fig:bycam}
\end{figure*}

\begin{figure*}
\centering
\includegraphics[width=\textwidth]{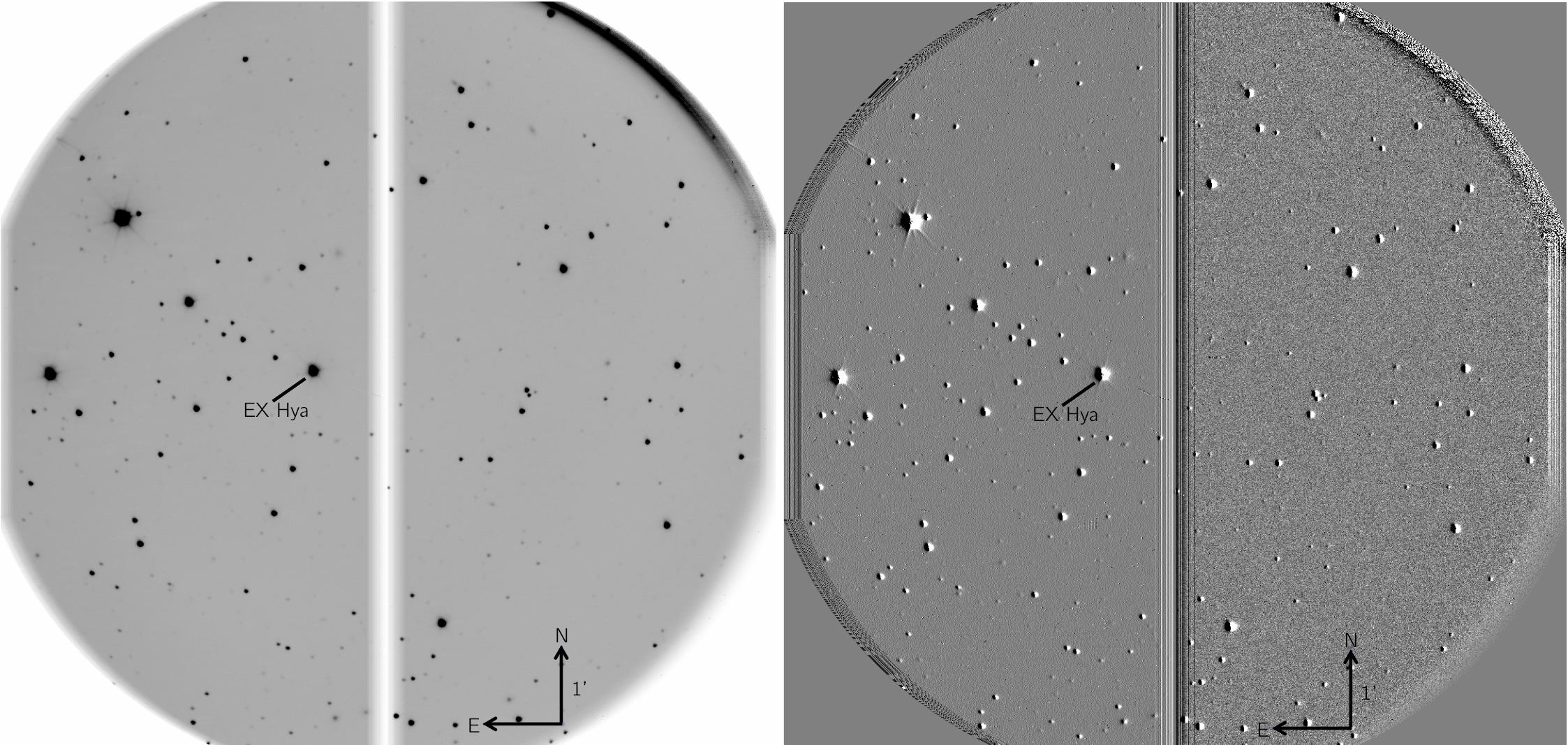}
\caption{{\it Left:} SALTICAM H$\alpha$ image of the IP EX Hya, made from a combination of 15 separate exposures taken on 2013 July 11, for a total exposure time of 1800s. North is up, East is to the left, and the length of the directional arrows corresponds to 1{\arcmin} on the figure. The full field of view is 8{\arcmin} in diameter, and the lighter stripe in the middle of the image is due to the gap between the two SALTICAM chips. The position of EX Hya is marked on the image, and there are no shells or shell fragments visible. {\it Right:} The unsharp masked image of the intermediate polar EX Hya, again processed using the same steps described in the caption for Figure \ref{fig:v1432aql}, also does not show any shells or shell fragments.}
\label{fig:exhya}
\end{figure*}

\begin{figure*}
\centering
\includegraphics[width=\textwidth]{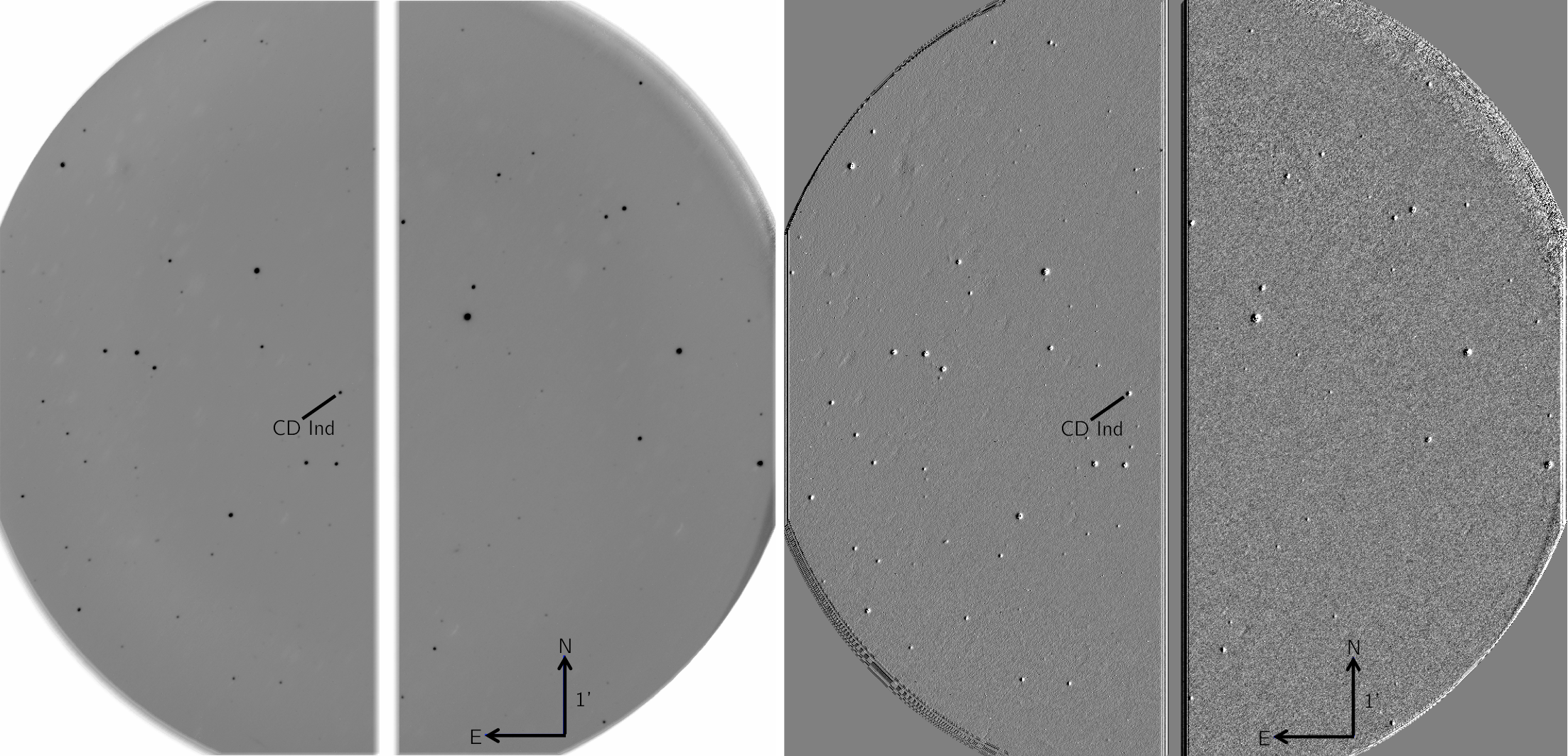}
\caption{{\it Left:} SALTICAM H$\alpha$ image of the AP CD Ind (RXJ 2115-58), made from a combination of 13 separate exposures taken on 2013 June 28, for a total exposure time of 1560s. North is up, East is to the left, and the length of the directional arrows corresponds to 1{\arcmin} on the figure. The full field of view is 8{\arcmin} in diameter, and the lighter stripe in the middle of the image is due to the gap between the two SALTICAM chips. CD Ind is marked on the figure, and no shells or shell fragments are visible in the image. {\it Right:} The unsharp masked image of CD Ind shown here was processed using the same steps described in the caption for Figure \ref{fig:v1432aql}. Again, it does not show any shells or shell fragments.}
\label{fig:cdind}
\end{figure*}

For all targets, the usable individual images were processed using the usual reduction steps in PyRAF and then summed using {\it imcombine} to obtain the greatest possible signal for each target. Compared to the observations from \cite{sahman2015a_arxiv}, we are able to go deeper on V1432 Aql and search a larger field of view for both V1432 Aql and BY Cam. The left-side images in Figures \ref{fig:v1432aql} to \ref{fig:cdind} show the stacked images for each system observed; no indications of any shells can be seen. To highlight any faint edges in the images and thus more thoroughly search for shells or shell fragments, we unsharp masked each combined image. The unsharpened versions are shown in the right side images of Figures \ref{fig:v1432aql} to \ref{fig:cdind}. None of the systems observed show visual evidence for a nova shell at any observed distance from the target star.

\section{Discussion}
\label{sec:discussion}

Following a nova eruption, the ejected shell will expand until it dissipates into the circumstellar medium. From Table 3 of \citet{pagnotta2014b}, which collects observed nova characteristics, the average expansion velocity of a classical nova is $1800$ km s$^{-1}$. Assuming this expansion velocity for any prior nova eruptions in the systems we observed, and considering the resolution of the detectors and the site conditions, we can calculate the size and age of all possible detectable shells. The distances used for each calculation are listed in Table \ref{tab:summary}. A certain number of years after an eruption, the shell will have expanded enough that it will be distinct from the image of the nova itself on the image (accounting for seeing, binning, and instrument effects), and first detectable in its smallest state. For each system, we assume the shell must be at least 5 pixels away from the star to be seen on the image; the minimum detectable shell sizes and ages are listed in Table \ref{tab:sizes}. For all of the systems we observed, shells from very recent novae can be expected to be seen, within the past two years for most of them. As the shell expands, eventually it will reach the edge of the field of view of the CCD. Since no dithering patterns were employed in our observations, we assume the total observed field corresponds to the full, unvignetted size of the instrument/chip, with the target system located at the centre. Using triangle geometry, we can calculate the physical shell size and thus the possible age, which then gives us a date back to which we can assume we may have seen a shell if the nova had erupted that recently and produced an observable shell.

For each system we calculate the shell sizes and ages assuming (a) constant expansion velocity, and (b) an expansion velocity that decelerates as described in \citet{duerbeck1987b}, which gives a mean half-lifetime of the velocities of 75 yr. From the Duerbeck paper, we can construct an exponential decay equation for the velocity over time:
\begin{equation}
v(t) = 5.68 \times 10^{10} \left( \frac{1}{2} \right) ^{\frac{t}{75}} \textrm{km yr}^{-1}.
\label{eqn:velocity}
\end{equation}
Taking the indefinite integral of Equation \ref{eqn:velocity} and solving for the integration constant given that $r(t=0)=0$ gives
\begin{equation}
r(t) = -6.15 \times 10^{12} \mathrm{e}^{-0.009\cdot t} + 6.15 \times 10^{12}  \textrm{  km}.
\label{eqn:radius}
\end{equation}
To calculate how long it would take the decelerating shell to reach the edge of our fields of view, we need time as a function of radius, so we rearrange Equation \ref{eqn:radius} to obtain
\begin{equation}
t(r) = \frac{-1}{0.009} \ln{\left(1-\frac{r}{6.15 \times 10^{12}}\right)} \textrm{  yr}.
\label{eqn:time}
\end{equation}
The amount of time necessary to observe the shells at their smallest sizes is the same for both cases (a) and (b), because it is such a short amount of time that no significant deceleration can have occurred on a level that we would be able to detect. Deceleration, however, does change the largest possible shell ages, increasing the amount of time we can expect to see the shell after the eruption, or essentially how far back in time we would be able to detect an eruption, because it takes longer for the shell to expand beyond the field of view if it is decelerating. If we take the \citet{duerbeck1987b} formulation at face value, we notice that the radius of the shell has an asymptote at $r=6.15 \times 10^{12}$ km. In some cases, the fields of view of our images are larger than this, so theoretically we could say that we would see all possible nova shells from an infinite amount of time in the past, however this is clearly unphysical, because we have observed at least two nova shells that are larger than the $r=6.15 \times 10^{12}$ km limit. AT Cnc and Z Cam, two dwarf novae with ancient nova shells \citep{shara2007a,shara2012b,shara2012c}, have shells with measured radii of $6.19 \times 10^{12}$ km and $2.15 \times 10^{13}$ km. The \citet{duerbeck1987b} result was empirically determined using observations of just four novae, so it is not altogether surprising that it is not universally applicable, but nevertheless it is a good first-order approximation of what we can expect, at least for the first 75 yr after a nova eruption. There is likely a strong dependence on the local circum- and inter-stellar medium, but measuring and modelling that is beyond the scope of this paper. For the cases in which our fields of view are larger than the $r=6.15 \times 10^{12}$ km asymptote, we can say only that we can see shells further back in time than in the no deceleration case, but cannot put a firm upper limit on the timeframe.

For V1432 Aql, we can rule out nova shells from eruptions that happened up to 118 or 145 yr ago in the constant expansion velocity case, depending on which distance measurement we use (187 or 230 pc, respectively; \citealp{ak2008a,barlow2006a}). Accounting for the deceleration of the shell, for both distances, we have cases where the field of view is larger than the asymptote, so we can say that 118 and 145 yr are lower limits.  For CD Ind, there are no shells detected from eruptions up to 59 yr ago in the constant velocity case, and 86 yr ago with a decelerating shell. EX Hya, the closest of our systems and the only non-AP, does not show shells from eruptions up to 35 yr ago or 43 yr ago, for constant and decelerating shell velocities, respectively. For BY Cam, with the caveat that the image is shallow, in the constant expansion velocity situation we rule out shells from eruptions that occurred as far back as 82 to 300 yr ago, again depending on the distance adopted (52 or 190 pc; \citealp{ak2008a,barlow2006a}). If the shell decelerates as described above, we can rule out shells from eruptions dating back to anywhere from 154 to $<$300 yr ago. These time constraints are listed for each system in Table \ref{tab:sizes}. 

Additionally, we were able to check whether the APs have large-scale ultraviolet-bright shells, similar to those found around Z Cam \citep{shara2007a} and AT Cancri \citep{shara2012b}. We searched the GALEX archive and found all but V1500 Cyg have images in both the FUV and NUV bands (135.0-175.0 nm and 175.0-280.0 nm, respectively). No shell is visible on any scale for any of the targets. The point spread function of GALEX is $\sim$6$\arcsec$ and confirms the H$\alpha$ non-detections on scales larger than this.

It is critical to remember that the lack of a nova shell does not equate to the lack of a nova eruption. \cite{wade1990} provides one of the first statistical looks at how many shells have been detected around classical novae, reporting that 26 of the approximately 200 known at the time had resolved remnants. \cite{downes2000a} did a survey of 30 recent, relatively nearby novae and found 14 shells using a combination of ground- and space-based imaging, giving a 47\% detection rate, although we note that this may be different from the actual shell formation rate. There are many reasons a shell may not be observed around a nova after its eruption even if it has formed: the amount of mass ejected might be so small that the shell density is low and the shell is undetectable even shortly after the eruption; or, enough time has passed since the eruption that, as the shell has expanded, its density and therefore surface brightness have decreased, making it too faint to detect; or, the shell might have expanded so quickly that it is larger than the field of view of the image.

There is another possibility for finding old nova eruptions in these systems: one can check through the major astronomical plate archives, namely those at the Harvard College Observatory in Cambridge, MA, and the Sonneberg Observatory in Sonneberg, Germany. There are scanning operations underway at both archives, which will allow for a quick check of the past behaviour of each of these objects once the fields are scanned and released. Although the Harvard operation, DASCH \citep{grindlay2012a}, has entered production scanning mode, it will likely be at least a few more years before all of the AP fields are scanned and available. Sonneberg is also scanning its plates, however they are not readily accessible offsite. Additionally, for a fully complete eruption search, it is recommended that the plates be examined by hand for evidence of eruption, especially in crowded fields, because it is possible that the eruption is only captured on one plate, and if for whatever reason that plate is not properly solved by the software pipeline, it will not be included in the digitized light curve results and the eruption will be missed. Although this method of searching for eruptions in archival plates only covers the last $\sim$120 yr, it is still a valuable resource in the attempt to find previous eruptions, and allows for the possibility of finding nova eruptions in systems that did not form detectable shells.

With no shell detections in our images, we cannot prove that any of the systems in our study\textemdash the three APs V1432 Aql, BY Cam, and CD Ind, and the IP EX Hya\textemdash had nova eruptions in the recent past that caused their asynchronicity today; we also cannot conclude that they did not have nova eruptions. It is possible that they erupted recently and the shells are fainter than our observations, for any of the possible reasons discussed above. In this case, deeper imaging, especially for BY Cam, is advised. It is also possible that the eruptions were further in the past than expected (i.e. our understanding of the models used to obtain $t_{\mathrm{sync}}$ are incorrect), especially for BY Cam, with its possible synchronization time of $>$3500 yr \citep{honeycutt2005a}, which indicates that wider fields of view are recommended. The obvious potential solution to both of these problems is the never-ending wish of the observational astronomer: more, better data. To quote directly from \citet{wade1990}, ``There are few branches of astrophysics where the old refrain, `More observations are needed!', is more applicable than to the study of resolved nebular remnants of classical novae." Deeper, higher-resolution exposures would allow us to search for fainter shells, and wider fields of view or well-designed dithering patterns would allow us to see a larger area around the central binary and therefore detect older, further nova shells, if they in fact exist. 

\section*{Acknowledgements}
This research was supported by the Kathryn W. Davis Postdoctoral Scholar programme, which is supported in part by the New York State Education Department and by the National Science Foundation under grant numbers DRL-1119444 and DUE-1340006.

This manuscript is based on observations made with the Southern African Large Telescope (SALT) under program 2013-1-AMNH-004 (PI: A. Pagnotta). We gratefully acknowledge that AMNH access to SALT is made possible by a generous donation from the late Paul Newman and the Newman Foundation.

This work is also based on observations obtained at the MDM Observatory, operated by Dartmouth College, Columbia University, Ohio State University, Ohio University, and the University of Michigan, using a filter borrowed from KPNO/NOAO, which was helpfully arranged by Eric Galayada at MDM. PyRAF is a product of the Space Telescope Science Institute, which is operated by AURA for NASA.

We thank Tom Maccarone for the initial discussion that sparked the idea for this project, Mike Shara for helpful discussions on the subject of nova shells, and Arlin Crotts for access to his MDM time. Jana Grcevich provided many useful suggestions throughout the course of this work, and Denise Revello provided invaluable coaching via the RAISE-W program; we are grateful to them both. The writing of this manuscript was continually accompanied by the dulcet tones of NPG, FH, the rest of the TBS Crew, and The Nixtape, which undoubtedly contributed to increased productivity.

\begin{table}
	\caption{The Asynchronous Polars}
	\label{tab:summary}
\begin{tabular}{lcccccc}
\hline
System & 		$P_\mathrm{orb}$ (h)		&	$P_\mathrm{spin}$ (h)	&	$\frac{P_\mathrm{orb} - P_\mathrm{spin}}{P_\mathrm{orb}}$	&	$\omega / \Omega $		&	$t_\mathrm{sync}$ (yr)	&	Distance (pc) \\
\hline

V1432 Aql	 &	3.3655	[1]\footnotemark[1]	&	3.3751	[2]	&	-0.0029	&	$-2.85 \times 10^{-3}$	[1]	&	110; $199^{+441}_{-75}$; $96.7 \pm 1.5$	[1; 3; 4]	&	187; 230	[5; 6]	\\
BY Cam	&	3.3544	[1]	&	3.3222	[2]	&	0.0096	&	$9.54 \times 10^{-3}$	[1]	&	1200; 1107; $\ge$3500; $250 \pm 20$	[7; 1; 8; 9]	&	52; 190	[5; 6]	\\
V1500 Cyg	&	3.3507	[1]	&	3.2917	[2]	&	0.0176	&	$1.76 \times 10^{-2}$	[1]	&	185; 150; 150-290	[7; 1; 9]	&	1038	[5]	\\
EX Hya\footnotemark[2]&	1.6376	[10]	&	1.1172	[11]	&	0.3178	&	\dots		&	\dots		&	56	[5]	\\
CD Ind	&	1.8467	[1]	&	1.8258	[2]	&	0.0113	&	$1.21 \times 10^{-2}$	[1]	&	\dots		&	94	[5]	\\
Paloma	&	2.6195	[12]	&	2.4328; 2.2709	[12]	&	0.0713; 0.1331	&	\dots		&	\dots		&	$\le 240$	[12]	\\

\hline
\end{tabular}
\end{table}

\footnotetext[1]{References: [1] \cite{campbell1999a}; [2] \cite{ramsay1999a}; [3] \cite{staubert2003a}; [4] \cite{andronov2007a}; [5] \cite{ak2008a}; [6] \cite{barlow2006a}; [7] \cite{piirola1994a}; [8] \cite{honeycutt2005a}; [9] \cite{pavlenko2013a}; [10] \cite{hellier1992a}; [11] \cite{hellier1996a}; [12] \cite{schwarz2007a}}
\footnotetext[2]{EX Hya is an IP.}

\begin{table}
	\caption{Shell Size \& Age Limits}
	\label{tab:sizes}
\begin{tabular}{lcccc}
\hline
System 	& 	Smallest Shell Size (\arcsec) 	&	Smallest Shell Age (yr)\footnotemark[3]	&	Largest Shell Age (yr)\footnotemark[3]	&	Largest Shell Age (yr)\footnotemark[4] \\
\hline

V1432 Aql\footnotemark[5]	&	3.57	&	1.8; 2.2	&	118; 145	&	$>$118; $>$145	\\
BY Cam\footnotemark[6]	&	4.48	&	0.6; 2.2	&	82; 300	&	154; $>$300	\\
EX Hya	&	4.91	&	0.7	&	35	&	43	\\
CD Ind	&	2.98	&	0.7	&	59	&	86	\\

\hline
\end{tabular}
\end{table}

\footnotetext[3]{Assuming a constant expansion velocity of the nova shell after the eruption.}	
\footnotetext[4]{Assuming a nova shell that decelerates after the eruption due to interactions with the circum- and inter-stellar medium. See Section \ref{sec:discussion} for more information on how the deceleration was calculated, as well as details on the "$>$" values in this column.}
\footnotetext[5]{V1432 Aql has two values in the Smallest and Largest Shell Age columns due to the two distances reported in the literature. The smaller values correspond to the 187 pc distance from \citet{ak2008a} and the larger to the 230 pc distance from \citet{barlow2006a}.}
\footnotetext[6]{BY Cam also has two values in the Smallest and Largest Shell Age columns due to two reported distances of 52 pc \citep{ak2008a} and 190 pc \citep{barlow2006a}.}

\bsp	
\label{lastpage}
\end{document}